\documentstyle[epsfig,prb,multicol,aps]{revtex}
\begin{document}

\title{Transport, optical and electronic properties of the half metal
CrO$_{2}$
}
\author{I.I. Mazin}
\address{Code 6691, Naval Research Laboratory, Washington, DC 20375 and CSI, George\\
Mason University, Fairfax, VA 22030}
\author{D.J. Singh}
\address{Code 6691, Naval Research Laboratory, Washington, DC 20375}
\author{Claudia Ambrosch-Draxl}
 \address{Institut f\"ur Theoretische Physik, Universit\"at Graz,
 Universit\"atsplatz 5, A-8010 Graz, Austria}

\date{\today}
\maketitle

\begin{abstract}
The electronic structure of CrO$_{2}$ is critically discussed in terms of
the relation of existing experimental data and well converged
LSDA and GGA calculations of the electronic
structure and transport properties of this half metal magnet,
 with a particular emphasis on optical properties. 
We find only moderate manifestations of many body effects.
Renormalization of the density of states is not large and 
is in the typical for transition metals range. 
We find substantial deviations from Drude behavior in the far-infrared optical
conductivity. These appear because of the unusually low energy of 
interband optical transitions. The calculated mass renormalization is
found to be rather 
sensitive to the exchange-correlation functional used and varies from
10\% (LSDA) to 90\% (GGA), using the latest specific heat
data. We also find that dressing of the
electrons by spin fluctuations, because of their high energy,
renormalizes
the interband optical transition at as high as 4 eV by
about 20\%.
Although we find no clear	
indications of strong correlations of the Hubbard type, strong 
electron-magnon scattering related to the half metallic band structure
is present and this leads to a nontrivial temperature dependence
of the resistivity and some renormalization of the electron spectra.
\end{abstract}
\begin{multicols}{2}

CrO$_{2}$ has been attracting substantial theoretical\cite
{schwartz,my,kub,nik,dG,allen,anis} and experimental\cite{sci,fuji,rho,dima}
interest in recent years,
in part because of its practical importance in
magnetic recording, and in part due to fundamental interest in
 its half-metallic electronic structure and the implications of
this.
In half metals, one spin channel is metallic and the other is
semiconducting. This
 results in unusual transport properties, including
anomalously large, but still 
metallic, resistivities at high temperature\cite{rho},
coupled with high mobilities at low temperature. Another consequence of
the half metal behavior is nontrivial magnetoresistance due to spin-dependent
tunneling\cite{shura,mrho}, with various potential applications.

 Despite a
large number of published band structure calculations, no 
full potential, all-electron calculations have been reported for 
CrO$_{2}$ except for calculations of Sorantin and Schwarz\cite{soren},
done in a study in chemical bonding of the rutile structure oxides.
The calculations of
Refs. \onlinecite{schwartz,my,kub,dG,anis} used atomic sphere approximations
for the crystal potential, while those of Ref. \onlinecite{nik} used a 
muffin-tin approximation.
Note that nonspherical part of the potential 
is of particular significance for a material with low site
symmetries like rutile CrO$_{2}$. The only nonspherical calculations 
reported so far are those of
Lewis {\it et al }\cite{allen}, however the accuracy of these 
calculations was limited by the use of a pseudopotential based
method.

  As was
observed early on\cite{my}, the exchange splitting is very sensitive to
computational details, hence the necessity of well-converged, all-electron, and
full-potential, state-of-the-art band structure 
calculations to establish the predictions of the LSDA which
may then be compared 
with experiment to clarify the possible role of many body effects
beyond the mean field.
Currently even such a basic quantity as density
of states at the Fermi level, $N(0)$, varies among different calculations
 from
0.69 states/Ry f.u. (Ref. \onlinecite{allen})
to 2.34 states/f.u.(Ref. \onlinecite{my}). Given that the  
reported experimental low temperature specific heat coefficients
range from 2.5 mJ/K$^{2}$(Ref.\onlinecite{fuji})
 to 7 mJ/K$^{2}$ (Ref.\onlinecite{gamma7}), such a
situation creates confusion
about the physical nature of this compound, including speculations
that strong correlations may lead to unusual renormalizations in CrO$_{2}$.

As mentioned, here we report well converged
full-potential all-electron  band structure calculations along with
 an analysis of the latest experimental
data. Our conclusion is that the existing experimental data are 
reconcilable with the calculated LSDA electronic structure without 
 additional strong correlation effects, but that strong electron-magnon scattering 
(as expected in half metals)
is presents and renormalizes the electron bands. With such renormalization 
the LSDA electronic structure suffice to understand reasonably well the body of experimental 
results.

The calculations were performed using the general potential linearized augmented
plane wave method (LAPW),
which is described in detail elsewhere\cite{OKA75,David}. 
Local orbital extensions were used to relax linearization errors and to treat
semicore states. Well
converged basis sets of over 600 functions for the six atom unit cell
were used.
 The radii of the muffin-tin spheres were 2 a.u. for Cr and 1.55
a.u. for O. The Hedin-Lundquist exchange-correlation potential was used for most
calculations (but see below).
The Brillouin zone averages during the iterations to
self consistency were performed with
105 inequivalent {\bf k}-points,
while the zone integrals for calculating transport
and related properties utilized
the tetrahedron method
with at least 770 inequivalent
{\bf k}-points. Convergence was carefully checked.

The resulting band
structure and electronic
density of states (DOS) are shown in Fig.\ref{FMbands} and \ref{FMDOS}.
The DOS agrees
well with that reported by Sorantin and Schwarz\cite{soren}.
First of all, let us note that the LSDA bandwidth $W$ is of the order
of 10 eV. This is to be compared with the Hubbard repulsion $U\approx 3$ eV
from constrained LSDA calculations\cite{anis}. The ratio of $U/W\approx 0.3$
is much smaller than in strongly
correlated oxides of Ni, Fe or Cu. 
One can expect at best a moderate effect
of electron-electron correlations beyond the mean-field LSDA result\cite{note1}.

In this
connection, it is instructive to look at the recent LSDA+U
calculations of Korotin 
{\it et al}\cite
{anis}. They compared their results with 
 an earlier LSDA band structure\cite{schwartz}, and
found some differences, mostly amounting to an upward 
shift of the unoccupied $d$-states  by $%
\approx 2$ eV. However the reported 
LSDA calculations deviate among themselves by
a comparable amount, and it worth trying to sort out the differences
genuinely related with the Hubbard $U$ from those related to the
computational details.

The results of Korotin {\it et al} \cite{anis} mainly differ from those of
Schwarz\cite{schwartz} in the position of the unoccupied spin-down
Cr-$d$ band. Compared to
the present LSDA results (Figs. \ref{FMbands},\ref{FMDOS})
this band is pushed up
by about 1.4 eV (more than twice less than $U$, due to the Cr-O hybridization\cite{anis1}). 
Thus the Fermi level appears to be
somewhat lower than in the middle of the gap, while in the LAPW calculations
reported here, as well as in Ref.\onlinecite{schwartz},
 it is closer to the top of the gap. The question is to what
extent this difference reflects real physics, that is, renormalization of
the bands due to electron-electron correlations beyond LSDA, and to what
extent it reflects the
difference in the computational technique and associated approximations.
 If the latter dominates,
full potential calculations are preferred
over  Atomic Sphere Approximation (ASA) calculations as
in Ref. \onlinecite{anis}. On the other hand, if the correlations beyond the
LSDA are very strong, the LDA+U calculations may be superior regardless
of the spherical approximation.
 It is worth
noting that although the statement of
Korotin {\it et al}\cite{anis}, that LSDA+U ``is
superior to LSDA since it can indeed yield a gap if the local Coulomb
interaction is large enough'', is often true,
 it is not so clear when the local interaction is
small compared to the band width, and the material is actually a metal. In
fact, when $U/W\sim 0.3,$ as in CrO$_2$,
one expects the crude way in which LSDA+U includes
correlations to be not better (if not worse) than the LSDA which
includes them roughly in a mean-field manner (and in any case the ASA then is a 
serious issue).

Recent photoemission investigations of
Tsujioka {\it et al}\cite{fuji}
found  large 
splittings of 5 eV between the main peaks in the occupied and empty Cr-$d$ 
densities of states. The authors compared that number with the corresponding
splitting in Ref. \onlinecite
{schwartz} ($\approx 1.5$ eV) and in Ref.
\onlinecite{anis} ($\approx 4.5$ eV), and interpreted that as an indication
of strong, important Coulomb correlations. However, this large difference
is not entirely due to the Coulomb correlations.
An immediate observation that can be made from
comparison of the results of Korotin {it et al}\cite{anis} and
Schwarz\cite{schwartz} is that
the Ref. \onlinecite{anis}
has not only the unoccupied $d$-band shifted up, but also
the weight inside the band redistributed. The latter effect can hardly
result from the Coulomb $U$ effect, because it is present also in our 
LAPW LSDA density of states (Fig. \ref{FMDOS}; note that splitting here is
about 2.5 eV).

In order to understand the remaining difference, and also to
explore the sensitivity of the exchange splitting to the technical
details, we performed a series of LMTO-ASA calculations, using first the
automatic maximum space filling atomic sphere
 radii, calculated from the Hartree
potential distribution by the Stuttgart LMTO ASA package\cite{TB}. Our
experience of working with this package is that this choice 
nearly always provides much better space filling and much smoother crystal
potentials than simple heuristics. In the case of CrO$_{2}$
this yielded $R_{Cr}=2.106,$ $R_{O}=1.872,$ $%
R_{E1}=1.728,$ and $R_{E2}=1.124$ a.u. Unlike the setup used in
Ref. \onlinecite{anis}
and other ASA calculations\cite{schwartz,my,usp}, there were 4
empty spheres of the 1st kind and 2 empty spheres of the 2nd kind per CrO$%
_{2}.$ The setup in Ref.\onlinecite{anis} generated too large sphere overlaps, which
according to the modern understanding of the LMTO\cite{TB,OPS} are above 
the maximum limit 
recommended for a second generation LMTO. However
we ran the program with that setup as well for the purpose
of comparison. The results are shown in Table \ref{thresh}. We find that this setup
consistently overestimates the exchange splitting by 0.2 eV. Another $\approx $%
1 eV then comes from  $U$, which may be considered as an estimate of 
 correlation effects beyond the LSDA in this system. It is worth noting that
the Generalized Gradient Approximation, 
discussed in more detail later in paper, takes correlations into account
in a more sophisticated, but still mean-field manner, and leads to an exchange
splitting about 0.3 eV larger than in LSDA, closer to LSDA+U. The gap in the 
spin-down channel, however, is the same in the GGA as in the LSDA, unlike LSDA+U. 

Let us now look at the electronic structure of the spin-up subband at the
Fermi level. Fig. \ref{FS} shows the calculated Fermi surface. The three bands
form a number of pockets; the main motif is a chain of alternating hole and
electron pockets extending along $k_z$ direction. The electron pocket
(``pseudocube'') is centered around $\Gamma ,$ and it touches a hole pocket
at the point {\bf k}$_{0}\approx (0,0,0.4\pi /c).$ These hole pockets extend
all the way up to the Z point where they touch the same pocket in the next
Brillouin zone. There are 0.12 electrons per formula unit, and the 
compensating amount of holes.
Interestingly, while the $\Gamma $-pocket is quasi-isotropic,
this hole pocket looks more like a pseudosphere with ``tentacles'' directed
towards the center of the $\Gamma $XMXZRAR prism\cite{points}%
, that is towards the point (%
$\pi /2a,\pi /2a,\pi /2c).$ The tentacle has a complex shape and 
extends from the
above mentioned point along (1\={1}0) direction
 hugging the pseudocube. The whole pocket resembles  a hammerhead
shark with one belly, four heads, and no tails. The main contribution to the
DOS comes from the ``heads''. Two additional small concentric electron
pockets occur between A and M. These two Fermi surfaces are degenerate on
the XMAR face of the Brillouin zone. The reason for that is the existence of
a twofold screw axis in the rutile structure; as was shown by Herring\cite
{her}, due to the time reversal symmetry, this leads to degeneracy of 
electronic bands on the plane perpendicular to the screw axis. This
degeneracy, generally
speaking, is lifted linearly in $k$, and the momentum matrix
elements between the bands are finite. This leads, in principle, to a finite
static interband conductivity\cite{IK},
$\sigma _{inter}(\omega \rightarrow 0)\neq 0$.
Since the length of the line where these two pockets touch is
small, this contribution is small, but still calls for additional caution
when analyzing experimental infrared conductivity in terms of a Drude model. 

The calculated DOS, as shown on Fig.\ref{FMDOS}, has a minimum
(``pseudogap''), but the Fermi level is not exactly at the minimum; the
calculated $N(0)=1.9$ states/eV cell corresponds to a linear
specific heat coefficient $\gamma (0)=2.24$ mJ/K$^{2}$ mole. This 
value is sensitive to the zone sampling and care was taken to ensure its 
convergence.
Tsujioka {\it et al}\cite{fuji}
have measured specific heat at low temperature and fitted
their data to the following expression: $\gamma (T)=2.5{\rm \ mJ/K}%
^{2}+0.0169{\rm mJ/K}^{4}T^{2}+1.36{\rm mJ/K}^{5/2}\sqrt{T},$ the last term
being a magnon contribution. It is worth noting that this curve, due to the
square root term, meets the origin with infinite slope, and already at $%
T=0.1$ K, for the parameters given in
Ref. \onlinecite{fuji}, $\gamma (T)\approx 3$
mJ/K$^{2}.$ This at least in part may explain the disagreement between various
authors about this value. The calculated $\gamma (0)$ is in close agreement
with the experiment, which is surprising (one expects substantial
renormalization from electron-magnon interactions), but less surprising
than the 
anomalous renormalization calculated by Lewis {\it et al}\cite{allen}. 
The number given by Tsujioka {\it et al}\cite{fuji}
may be an underestimate, given the
sharp change of $\gamma (T)$ near $T=0.$ We, however, think that this number
is more realistic than the 2--3 times larger values quoted in Refs. \onlinecite
{gamma7,gamma5}. On the other hand,
the above mentioned minimum in $N(E)$ occurs
at only 30 meV above the Fermi level. The strong variation in this region
and high sensitivity of the exchange splitting to the computational parameters
together lead to an extraordinary sensitivity of the calculated $N(0)$ to the
exchange-correlation potential (particularly, to gradient
correction). In fact, in GGA calculations
the Fermi level falls exactly at the minimum and the calculated DOS is 
only 0.95 states/eV cell, half of the LSDA value. 
Depending on the exchange-correlation functional used, the calculated
bare electron
specific heat coefficient can be anything from 1.1 to 2.3 mJ/K$^{2}$ mole,
and correspondingly the needed mass renormalization anything from 2.5 to 1.1	.

The pseudogap in the DOS also leads to a minimum in the plasma frequency
as a function of the Fermi energy position.
At the actual position of the
Fermi level, the plasma frequency is calculated to be 2.1 eV\cite{plasma}.
This is smaller than the value (3.35 eV) derived by Basov {\it et al}\cite
{dima} by integrating the measured optical conductivity up to 12 000 cm$^{-1}
$ (1.49 eV). The reason is that, as mentioned above, there are
interband transitions with vanishing energy in this system, and a large
part of the spectral weight integrated up to 1.5 eV comes from these
interband transitions. Note that (Fig.\ref{FMDOS}) the plasma frequency is
much less sensitive to the position of the Fermi level than the DOS ---
changes in the value of $N(E)$ are compensated by the opposite changes
in the Fermi velocity.
In the above 
mentioned GGA calculations, despite a drastic change in $N(0)$, $
\Omega_p$ was only 5 and 7\% smaller for the $xx$ and $zz$ components,
 respectively, than in LSDA.

Combined with the small residual resistivity (20 $\mu
\Omega $ cm in static measurements\cite{rho} and 10 $\mu \Omega $ cm or less
when extrapolating the optical conductivity to zero\cite{dima}), the
calculated $\omega _{p}$ calls for an unusually long relaxation time at low
temperature: $\gamma =\hbar /\tau =50$ to 100 cm$^{-1}$ (the first number
is based on DC resistivity and the second on the optical data). Basov {\it et al}\cite
{dima} fitted their low-frequency data ($\omega \alt150$ cm$^{-1})$ by a
Drude formula with $\omega _{p}\approx 1.7$ eV and $\gamma \approx 30$ cm$%
^{-1}.$ So the LSDA calculations again fall within the range of the
the different experiments.

Another characteristic that can be extracted from the
calculations is the average Fermi velocity. Note that one cannot
reliably extract this number from the optical experiments,
since the required effective mass is
unknown. Although the ratio of squared plasma frequency and the DOS,
formally speaking, gives the average squared Fermi velocity, this procedure
becomes misleading when the effective mass is rapidly changing over the
Fermi surface: it overweights heavy-mass parts of the FS and reduces the
calculated $\langle v_{F}^{2}\rangle ,$ while the actual transport is dominated
by the lighter carriers with higher velocity. However, even for the lightest
band at the Fermi level the calculations 
give $\sim 0.2\times $10$^{8}$ cm/s. The
estimation of Ref.\onlinecite{dima} (0.6-0.9$\times $10$^{8}$ cm/s) is
substantially higher; we ascribe this partially to the
neglect of the effect of interband transitions and
the deviation of the carrier mass from the free electron mass (and indeed
10$^{8}$
cm/s is an unusually large number for a $d$-band metal). Correspondingly, their
estimate for the mean free path at $T=10$ K should be reduced from 1500 to
300 \AA , which is not extraordinary large. The former number is hard to
explain by freezing out of spin-flip impurity scattering
 (due to the half-metallic character of CrO$_2$),
as it is hard to
imagine that all charge defects are absent and the only possible scattering
channels are associated with the spin flip. 

On the contrary, the temperature dependence of the resistivity indicates
a significant
 role played by spin scattering at finite temperature. As shown in Ref. 
\onlinecite{rho}, the resistivity below the room temperature can be fitted by a
quadratic dependence, or slightly better with a power law with an exponent $%
n=2.3-2.5.$ Various models for the scattering by spin fluctuations give
exponents between 2 and 3.5, but in any event much larger that one.
Interestingly, $\rho (T)$ for CrO$_{2}$ looks very similar to that of 
other so-called ``bad metals''. A typical example is SrRuO$%
_{3}$, with $\rho (T)$ closely reminding $\rho (T)$ in CrO$_2$,
including the change of the slope at the Curie
temperature\cite{SRO3}. Interestingly, in Ref. \onlinecite{SRO3} 
the observed resistivity was argued to be {\it not} related
 to magnon scattering because the prefactor
of $T^{2}$ seemed too large ($\sim 10^{-8}\Omega $ cm/K$^{2})$ compared to
the corresponding term for elemental ferromagnets
($\sim 10^{-11}\Omega $ cm/K$%
^{2}).$ As we discussed elsewhere\cite{ruthen}, the origin is coupling
of magnetism and covalent bonding via O Stoner factor. The physics of CrO$_2$,
despite its being also a bad metal, is very different. What these
two materials (and probably other bad metals as well) have in common is
that in both metals changing magnetic
order has strong qualitative effects on the band structure, and
because of that electrons are scattered
anomalously strongly by spin fluctuations (in this case electron-magnon
interaction is due to what can be called
``magneto-covalent'' coupling). Note that
the $T^{2}$ prefactor in CrO$_{2}$ is of the same order as in SrRuO$_{3}$
(about three times smaller).

Optical spectra of CrO$_{2}$ reveal
 a number of interesting features\cite{dima}. Two
features observed in the high-energy part of the spectrum are a peak at $%
\approx 3$ eV, easily identifiable with transitions across the spin-minority
gap, first pointed out by Uspenski {\it et al}.\cite{usp}, and a weaker peak
(essentially, a shoulder) at $\approx 2$ eV. This latter feature coincides
both in position and relative strength with a peak calculated in
Ref.\onlinecite{usp}
in the polarization {\bf E}$\parallel c,$ coming from the spin-majority
band (this feature is weaker in the other polarization\cite{dimap}). The
third observed feature is a broad maximum between 500 and 10000 cm$^{-1}$
(0.06 to 1.2 eV), interpreted in Ref.\onlinecite{dima} as intraband conductivity
shaped in a non-Drude way by a frequency-dependent relaxation time. However,
a closer examination of the results of Uspenski {\it et al}.\cite{usp} reveal
an interband feature starting at about 0.1 eV and continuing to 0.9-1 eV, at
least in the {\bf E}$\perp c$ polarization. 
The relative strength of these three features in the calculations
may be questioned,
particularly the excessive weight in the
lowest one. We should keep in
mind that a substantially greater error is expected in the calculation of
the matrix elements within LMTO-ASA technique, than in the calculations of
the band structure itself\cite{ZfP}. Besides, there is the 
substantial polarization dependence revealed by the calculations. This
introduces additional uncertainty in the Kramers-Kronig procedure\cite{note2}%
. 

Uspenski {\it et al}.\cite{usp} did not discuss the low
frequency ($\omega <500$ cm$^{-1})$ region in detail.
However, following the work
of Basov {\it et al}\cite{dima}, one may look at this region
more closely. This requires knowledge of the band structure near the Fermi
level with a better accuracy than that provided by the LMTO-ASA method.
Besides, LMTO-ASA introduces uncontrollable errors into optical matrix elements,
hence the need for full potential optical calculations.
We have performed such calculations using the Wien LAPW code with the 
optical extension, described elsewhere\cite{Claudia}. The band structure
calculated with this code differs from the one calculated using the
LAPW code\cite{David} used in the transport calculations by less than 1 mRy. 

As mentioned, CrO$_{2}$ is special in the sense that it
has non-zero interband conductivity at $\omega \rightarrow 0.$
To demonstrate this, we show in  Fig.\ref{JDOS} the 
calculated joint density of states (JDOS), defined as
\end{multicols}
\rule[10pt]{0.45\columnwidth}{.1pt}
\begin{equation}
J_{ij}(\omega )=\sum_{{\bf k}}f(\varepsilon _{{\bf k}i})f(-\varepsilon _{%
{\bf k}j})\delta (\varepsilon _{{\bf k}j}-\varepsilon _{{\bf k}i}-\hbar
\omega ),
\end{equation}
which is proportional to the optical conductivity due to the $i\rightarrow j$
interband transitions in the constant matrix element approximation:
\begin{equation}
\tilde{\sigma}_{ij}(\omega )=\frac{\pi }{\omega V}\sum_{{\bf k}%
}p_{ij}^{2}f(\varepsilon _{{\bf k}i})f(-\varepsilon _{{\bf k}j})\delta
(\varepsilon _{{\bf k}j}-\varepsilon _{{\bf k}i}-\hbar \omega )=\frac{\pi
p_{ij}^{2}}{\omega V}J_{ij}(\omega ).
\end{equation}
\begin{flushright}\rule{0.45\columnwidth}{.1pt} \end{flushright}
\begin{multicols}{2}
Here $V$ is the volume of the unit cell, and $p_{ij}$ is the dipole matrix
element for given polarization. As one can see from Fig.\ref{JDOS}, there is
indeed a linear in $\omega $ component. A closer look uncovers that there is
also a quadratic in $\omega $ component in JDOS. This comes from the
transitions around the point (0,0,0.4$\pi /c)$ where two bands touch (the
``pseudocube'' and the belly of the ``shark'').  Even more
interesting is the steep increase of the JDOS above 700 cm$^{-1}.$ This is the
threshold for the transitions from the ``hammers'' into the ``pseudocube''.
The weak
 dispersion in one direction of the band forming the ``hammers'' provides
a large phase space for this transition.

The calculated optical conductivity including matrix elements is shown
in Fig.\ref{sigmaall}. In the following discussion we shall concentrate
at the low energy ($\omega<4.5$ eV) region, studied in Ref.\onlinecite{dima}.
This region is blown up on 
 Fig.\ref{IRboth}, where we also show the experimental
data from Fig. 2 of the Ref.\onlinecite{dima}. We have added a Drude part
corresponding to the $\sigma_0=6000$ ($\Omega$ cm)$^{-1}$ and the calculated 
plasma frequency of 2.1 eV. First of all, we observe that all essential
features visible in the experiment are reproduced: a narrow intraband
peak with no deviation from Drude shape, interband transitions within the
$t_{2g}$ manifold, starting from $\omega=0$ and extending up to 1.5 eV 
(1.3 in the experiment), followed by a wide absorption band between 2 and 4.8
eV (between 1.7 and 4.5 in the experiment), which is associated 
with transitions between the bonding and antibonding $e_g$ bands.
The $e_g$ manifold includes some dispersionless bands (which are understood
in the tight binding model described in Ref.\onlinecite{anis}). These 
in turn result in an absorption peak near 2.5 eV. It is tempting to
associate this peak with a shoulder at 2 eV, found in the experiment,
although by doing so we risk indulging in overinterpretation. Nonetheless,
in this case the broad hump in the experimental curve, centered
at 3 eV, may be associated with the feature in the calculated $\sigma
^{zz}$ at 3.5-4 eV. It is notable that the experimental features systematically
occur at frequencies 10-20\% lower than the corresponding
 calculated ones. This must
be due to many body renormalization effects. Contrary to the
assumption of Refs.\onlinecite{anis} and
\onlinecite{fuji}, these renormalizations
seem to be related with the exchange of virtual magnons
rather than with the Hubbard repulsion, as this would force apart occupied
and unoccupied states and would shift optical transitions to higher energies.
(cf. Eliashberg renormalization due to electron-phonon interaction;
however, because of the high frequency of
the spin excitations, this effect is seen even far away from the Fermi level).
Again, one should exercise caution here, because the crystalline anisotropy,
barely noticeable in the intraband conductivity, increases with frequencies
(as one can see from Fig.\ref{IRboth}). First of all, we do not know with
which weights the two polarizations appear in the experimental curve.
Second of all, even if we knew that, and would have averaged the two
curves in Fig.\ref{IRboth}, this procedure would not have produced the
same results as averaging two reflectivities and restoring the conductivity
from the average reflectivity (which would correspond to the 
experimental procedure). To illustrate that, we compare the average 
calculated reflectivity (assuming 1:1 weighting of $R_x$ and $R_z$,
that is, an orientation preferential for $z$ polarization) with 
the reflectivity calculated from the average dielectric function
(Fig. \ref{R}).
While in general the curves are similar, the latter overemphasizes
the feature at 1.3 eV, right below the screened plasma edge at 1.6 eV.
Indeed, we observe from Fig.\ref{IRboth} that the structure in the
experimental curve in this energy region is substantially weaker than 
in the calculated one.

The last note regarding comparison with the
experiments regards the low-temperature conductivity displayed in Fig. 4 
of the Ref.\onlinecite{dima}.  
As we have noted before, the Fermiology of CrO$_{2}$ is unusually sensitive
to the approximations used for the exchange-correlation functional. In
particularly, the above mentioned smaller density of states in GGA
is due to the fact that the hammerheads of the hole pockets entirely 
disappear when the Fermi level is raised by 30 meV.
So, one should take the calculation
in the far infrared ($\omega\alt 500$ cm$^{-1}\approx 60$ meV)
region with a grain
of salt. With this in mind, the agreement between the frequency of the
threshold (700 cm$^{-1})$ seen in Fig.\ref{JDOS}
 and the frequency $E_{3}\approx 300$ cm$^{-1},$
where the experimentally determined low temperature optical conductivity starts
to deviate from the Drude law, is very reasonable. Note that the original
interpretation of Basov {\it et al} was in terms of {\it intra}band
transitions and a strongly frequency-dependent relaxation time. They assumed
a threshold in the relaxation time around 500 cm$^{-1}$ and
 ascribed this threshold to the minimum energy for spin-flip
scattering. If the main scattering mechanism is by magnetic impurities, then
indeed it is easy to show that in the constant scattering amplitude
approximation the relaxation time depends on frequency as 
\end{multicols}
\rule[10pt]{0.45\columnwidth}{.1pt}
\begin{equation}
1/\tau (\omega )\propto \int_{-\omega /2}^{\omega /2}\left[ N_{\downarrow
}(\varepsilon -\omega /2)N_{\uparrow }(\varepsilon +\omega /2)+N_{\downarrow
}(\varepsilon +\omega /2)N_{\uparrow }(\varepsilon -\omega /2)\right]
d\varepsilon ,
\end{equation}
\begin{flushright}\rule{0.45\columnwidth}{.1pt} \end{flushright}
\begin{multicols}{2}
and obviously is zero below the spin-flip transition threshold. However
this interpretation relies on a threshold of 500, or at most 1000 cm$^{-1},$
while our LSDA calculations give at least 0.25 eV, that is
2000 cm$^{-1}.$ This number is defined by the exchange splitting and
the many body correlation effects neglected in the LSDA are likely
to increase it even further
(cf. Ref.\onlinecite{anis}).
Interpretation in terms of interband transitions fits
the experimental picture much better.

The level of 
agreement between the observed low-frequency plasma energy, 1.7 eV, and the
calculated value of 2.1 eV is satisfactory. It is tempting to ascribe
overestimation of the plasma frequency  to many body effects (cf. high-$T_{c}
$ cuprates where the LSDA underestimates $\omega_p$ by a factor of two).
However, one should be very cautious about such an analogy,
because renormalization in cuprates affects the
quasiparticle velocity but not the area of the Fermi surface, so that the
DOS ($\propto \int dS_{F}/v_{F})$ is underestimated by about as much as $%
\omega _{p}^{2}$ ($\propto \int v_{F}dS_{F})$ is overestimated. Here, however,
both $N(0)$ and $\omega _{p}^{2}$ seem to be overestimated, suggesting that it
is the total area of the Fermi surface that is overestimated. Unlike the high
$T_c$ cuprates, the Luttinger theorem does not fix the size of the Fermi
surface because both electron and hole sections
 are present in CrO$_2$. The most
natural explanation of the overestimated plasma frequency
in the LSDA is thus  not a failure of the one-electron picture,
but inaccuracy of the density functional
 approximation used in the calculations. The fact that GGA
calculations yield  smaller
(substantially) $N(0)$ and (somewhat) $\omega_p$ illustrates this point.
We should emphasize that we do not claim the GGA is necessarily a better
approximation than LSDA for CrO$_2$; it does give larger separation 
between the occupied and unoccupied $d$ states, both in the spin-flip
and in the same spin channels, thus being closer to LDA+U than LDA is.
We are, however, sceptical about superiority of LDA+U in case of
CrO$_2$, and indeed the optical data suggest that the LSDA bands should be
corrected in the direction opposite to LDA+U or GGA, that is, the band
widths should be smaller, as well as the separation between occupied and 
unoccupied bands.  

Finally, an interesting question arises about the evolution of the optical
and transport properties above the Curie temperature. One may expect that at
sufficiently high temperatures the optical response would be determined by
the paramagnetic bands structure rather than by the ferromagnetic one, but
this temperature may be much 
higher than Curie temperature. In fact, local moment systems% 
, {\it e.g.} Fe, retain magnetic splitting at the
atomic level well above the Curie temperature where the spins lose their long
range order. One way to detect such a situation in zero temperature
calculations is to compare ferro- and antiferromagnetic energies
and moment. If the
difference between the two (set by the interatomic exchange energy) is small
compared to the intraatomic exchange (Hund energy) taken as the difference
between a magnetic and the  nonmagnetic state, and the moments are not
sensitive to the local order,  one has a local
moment system.

CrO$_{2}$ is an intermediate case: the
antiferromagnetic structure
appears higher in energy than the ferromagnetic by about a quarter of an
eV/cell (which corresponds to $J\sim 200$ K, in accord with the experiment),
 while the one without spin polarization is higher again in energy than the
antiferromagnetic one by about the same amount\cite{note3}. On the other
hand, the antiferromagnetic CrO$_{2}$ is metallic (not a half-metal),
but the
magnetic moment inside the Cr atomic sphere is only weakly
 changed (1.56 $\mu_B$ {\it vs.} 1.97 $\mu_B$).
%As such, the paramagnetic band structure may  be a
%reasonable, although not perfect,
% starting point for interpreting optics at temperatures higher
%than $T_{C},$ $T\agt T_{C}$ (of course, strong scattering by magnetic
%fluctuations makes Fermi liquid picture in general less reliable at such
%temperatures). With this in mind, we show paramagnetic band structure and 
%density of states in Figs.\ref{NMbands} and \ref{NMDOS}. Note that the
%DOS at the Fermi level and the plasma frequency are larger than in the
%ferromagnetic state: $N(0)=$4.2 states/eV spin f.u., $\omega_{px}=$4.5
%  eV, and  $\omega_{pz}=3.6$ eV.

To summarize, we present the
first all-electron and full potential LSDA
band structure and optical coefficients
 calculations for  CrO$_{2}$. We show that apart from anomalously
strong electron-magnon scattering, which naturally
follows from its half-metallic
band structure,
the experimental observations may be interpretable in terms
of quasi-one-electron band theory. Thus there is no experimental
smoking gun in regard to strong correlations related exotic phenomena
in CrO$_2$. Contrary to some conclusions deduced from 
less accurate band structure calculations, we find very modest specific
heat renormalization, and the
optical plasma frequency in reasonable agreement with 
the low temperature far infrared optical experiments, and interband transitions
starting from zero frequency. The latter explains the non-Drude shape of the
observed optical conductivity in the range of 500 to 4000 cm$^{-1}$.

The authors are grateful to D.N. Basov for numerous stimulating discussions
and for making the results of Ref.\onlinecite{dima} available prior to
publication. This work was supported in part by the Office of Naval Research.

\setlength{\columnwidth}{3.5in}
\nopagebreak
\begin{figure}[tbp]
\centerline{\epsfig{file=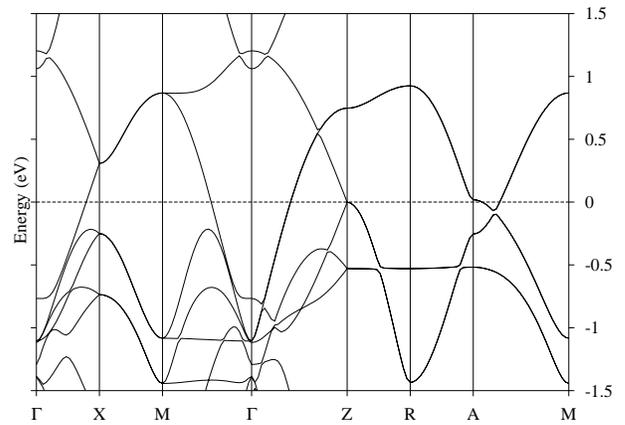,width=0.95\columnwidth}}
\vspace{0.1in}
\nopagebreak
\caption{ Calculated band structure of ferromagnetic CrO$_2$ in the spin-majority
channel. The Fermi energy is
denoted by the dashed horizontal line at 0.}
\label{FMbands}
\end{figure}

\begin{figure}[tbp]
\centerline{\epsfig{file=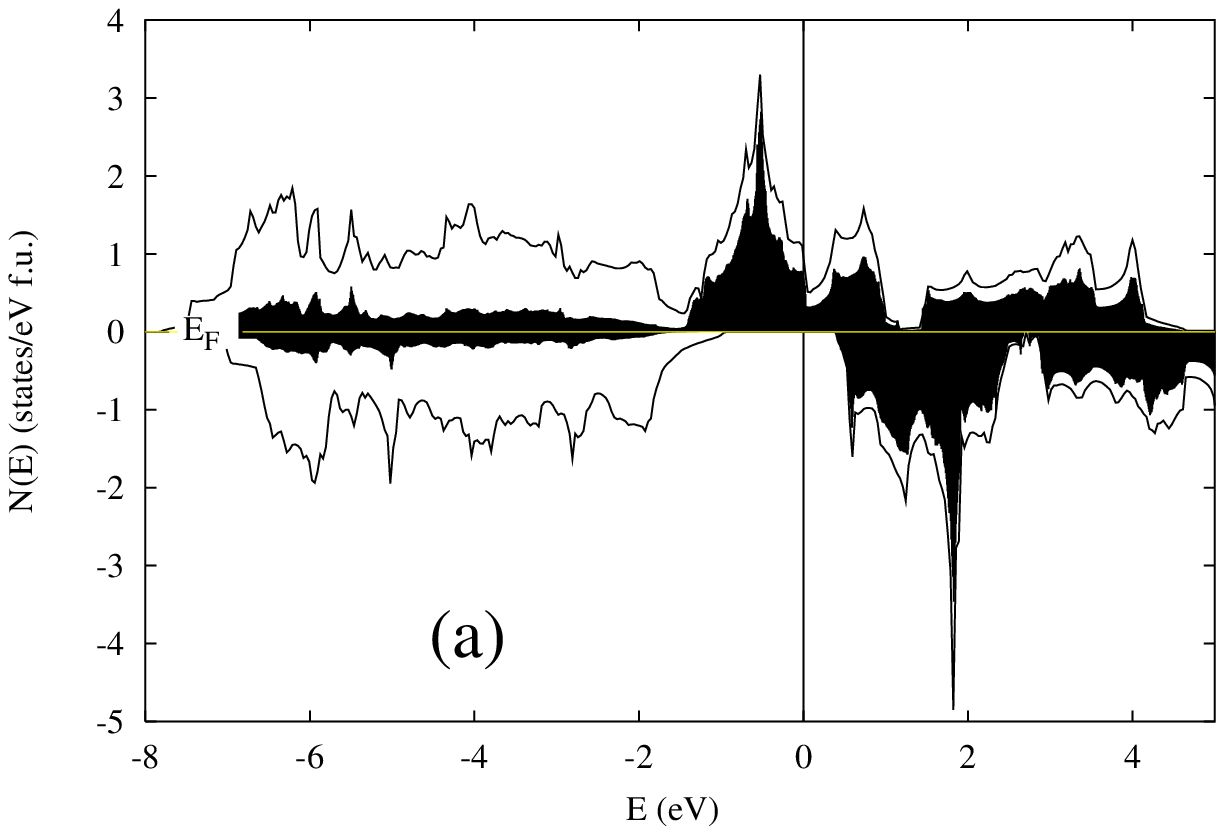,width=0.95\columnwidth}}
\centerline{\epsfig{file=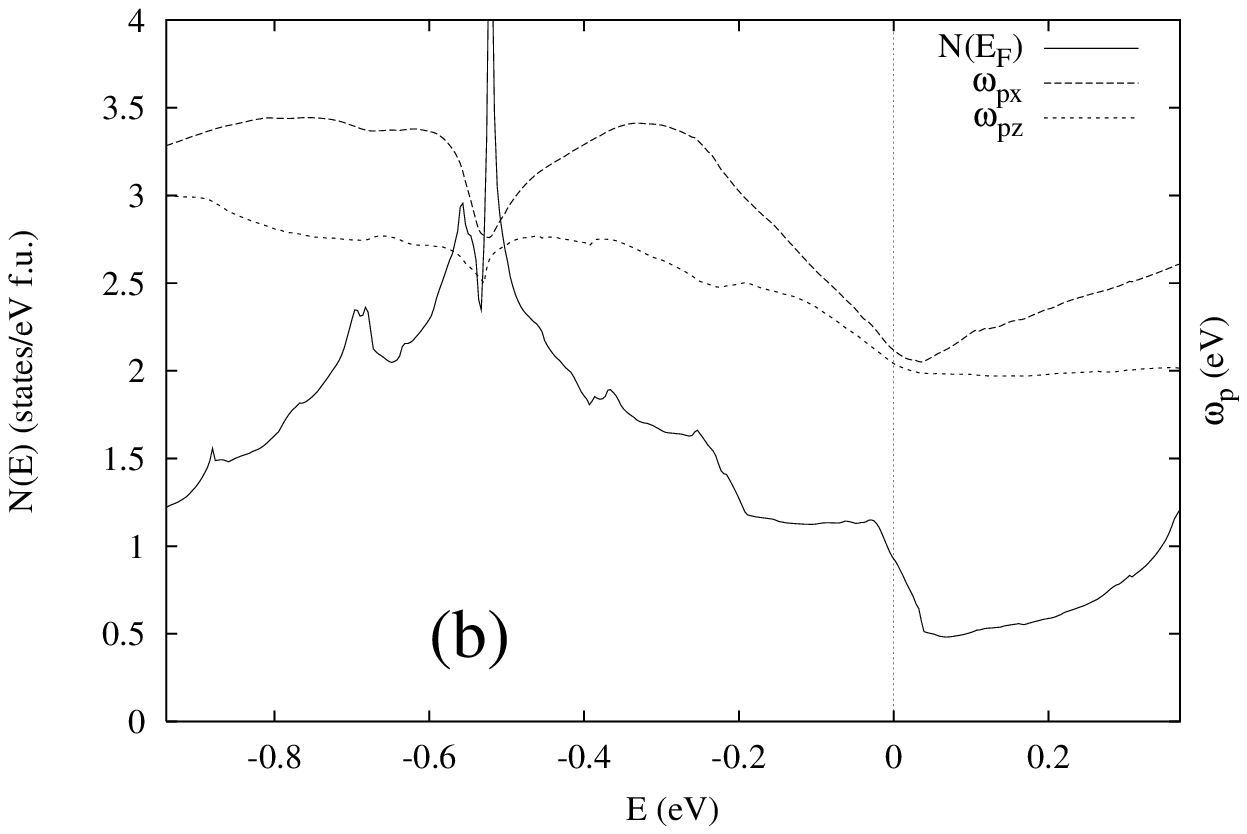,width=0.95\columnwidth}}
\vspace{0.1in} 
\nopagebreak
\caption{(a) LAPW DOS of ferromagnetic CrO$_2$ on a per formula unit basis.
The shaded area is the Cr $d$ contribution.
 The Fermi energy is at 0. (b) Blowup of the DOS for spin-majority
channel inside the spin-minority gap. Also shown are plasma frequencies
for the two crystallographic directions as function of the chemical potential.}
\label{FMDOS}
\end{figure}

\begin{figure}[tbp]
\centerline{\epsfig{file=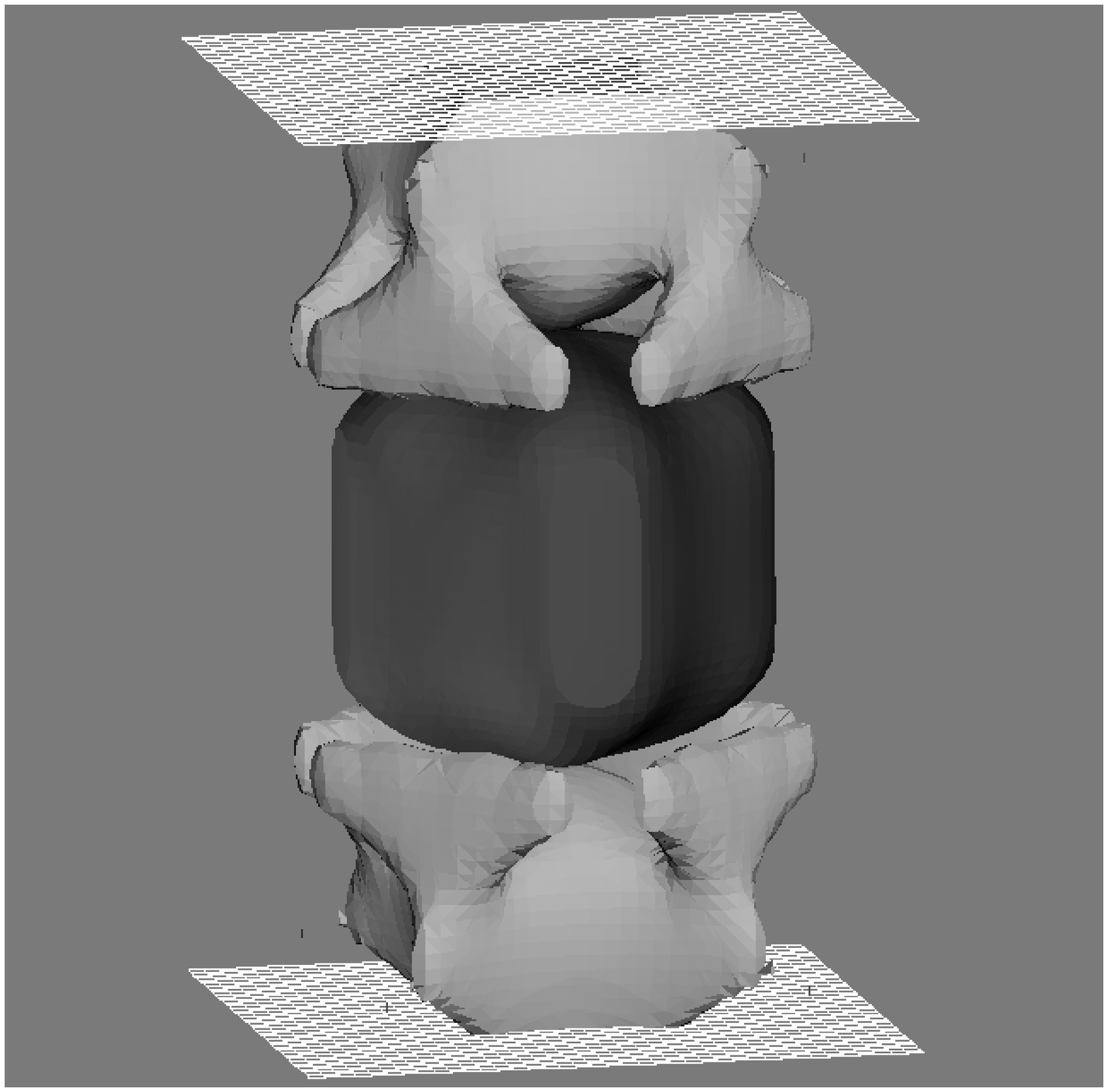,width=0.95\columnwidth}}
\vspace{0.1in}
\nopagebreak
\caption{LAPW Fermi surface of CrO$_2$. Note  the electron ``pseudocube''
around the $\Gamma$ point and hole ``hammerheads'' touching the 
``pseudocube'' between  $\Gamma$ and Z. Small pockets along the AM line.}
\label{FS}
\end{figure}

\begin{figure}[tbp]
\centerline{\epsfig{file=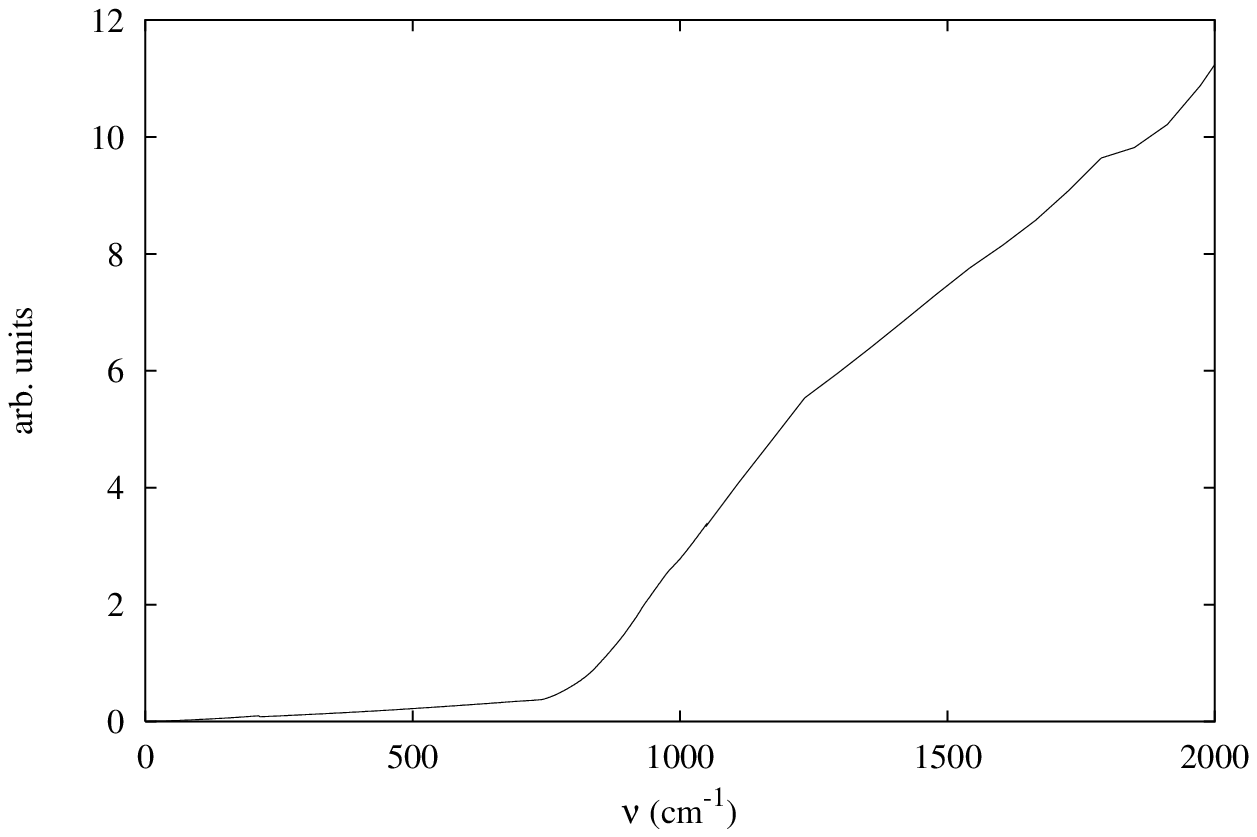,width=0.95\columnwidth}}
\vspace{0.1in}
\nopagebreak
\caption{Joint DOS of ferromagnetic CrO$_2$ in the far infrared
region}
\label{JDOS}
\end{figure}

\begin{figure}[tbp]
\centerline{\epsfig{file=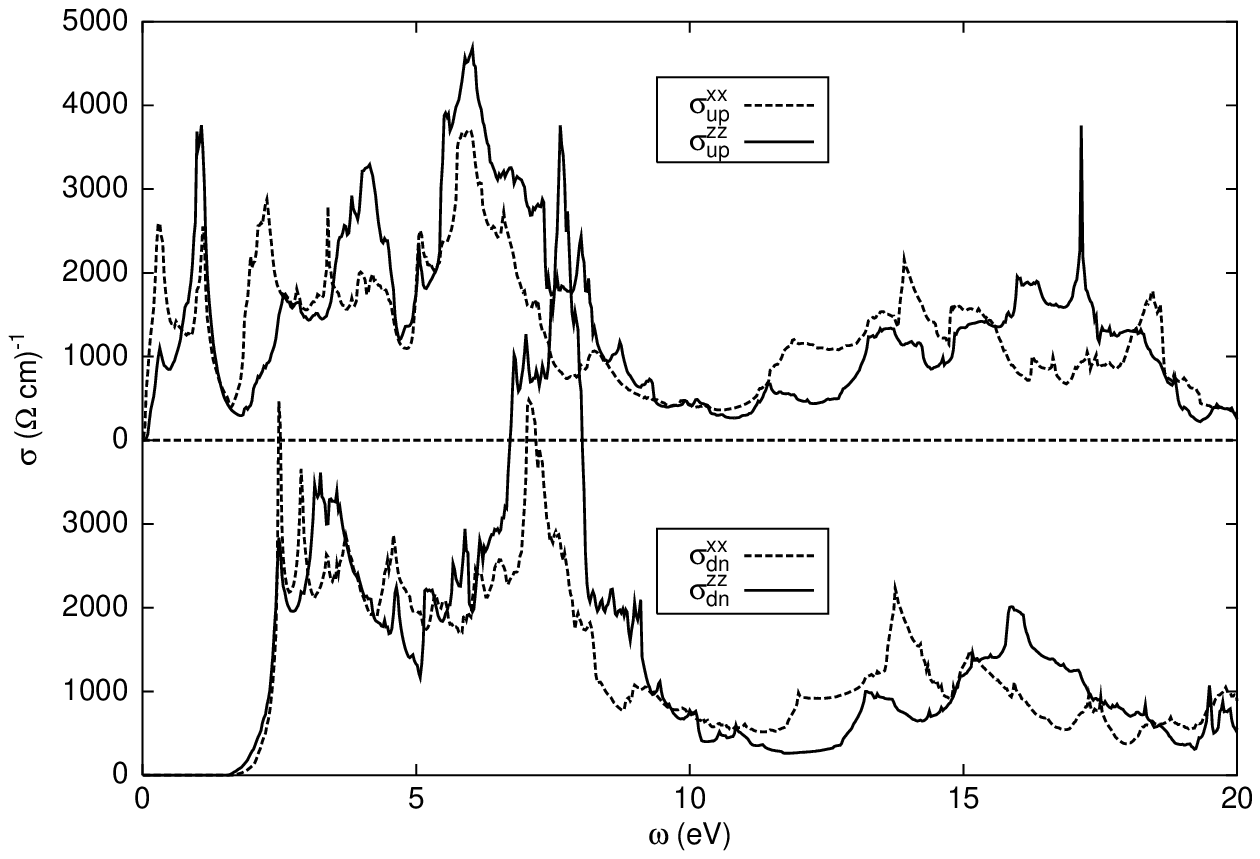,width=0.95\columnwidth}}
\vspace{0.1in}
\nopagebreak
\caption{Calculated optical conductivity
of CrO$_2$. Spin-up and spin-down channels are shown separately.}
\label{sigmaall}
\end{figure}

\begin{figure}[tbp]
\centerline{\epsfig{file=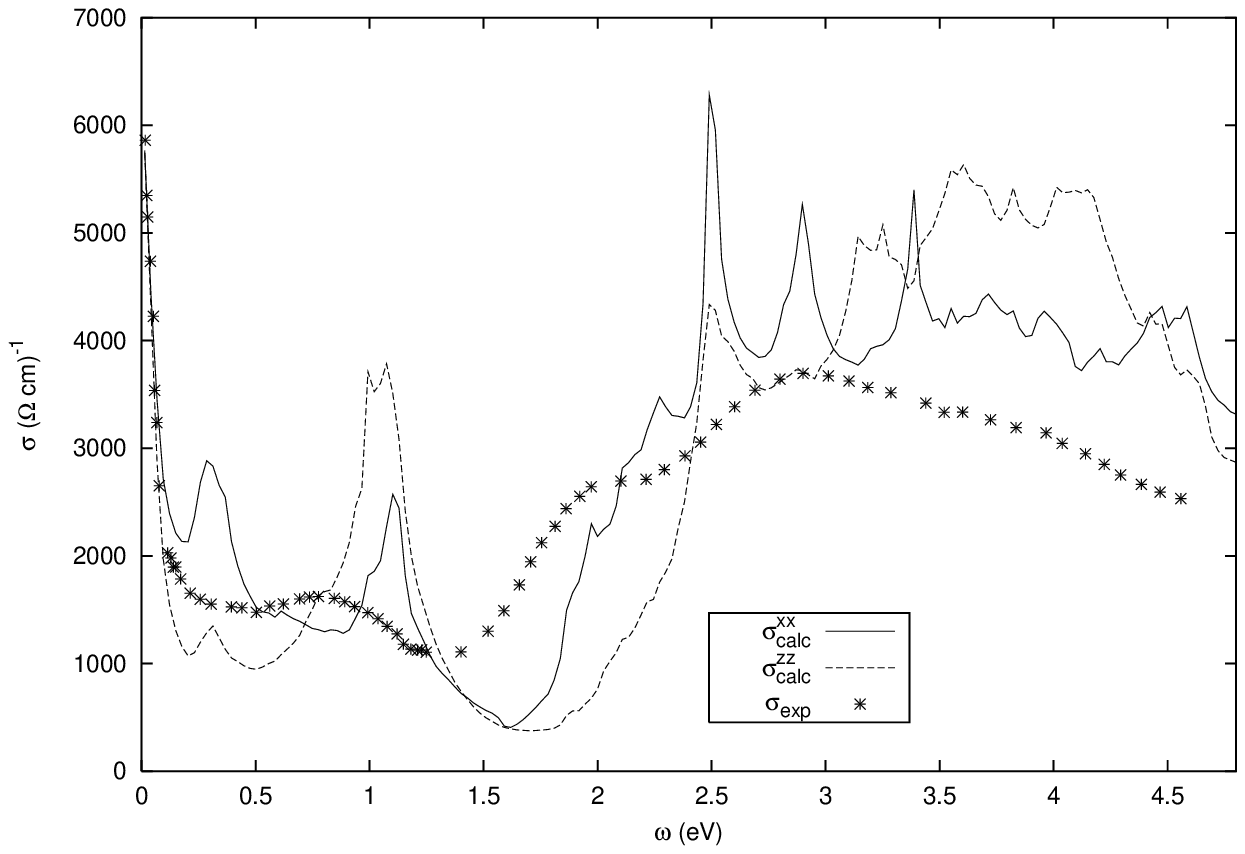,width=0.95\columnwidth}}
\vspace{0.1in}
\nopagebreak
\caption{Calculated optical conductivity of CrO$_2$.
Spin-up and spin-down channels are added together. Symbols show
experimental data at 300 K from Ref.\protect\onlinecite{dima}, Fig. 2.}
\label{IRboth}
\end{figure}

\begin{figure}[tbp]
\centerline{\epsfig{file=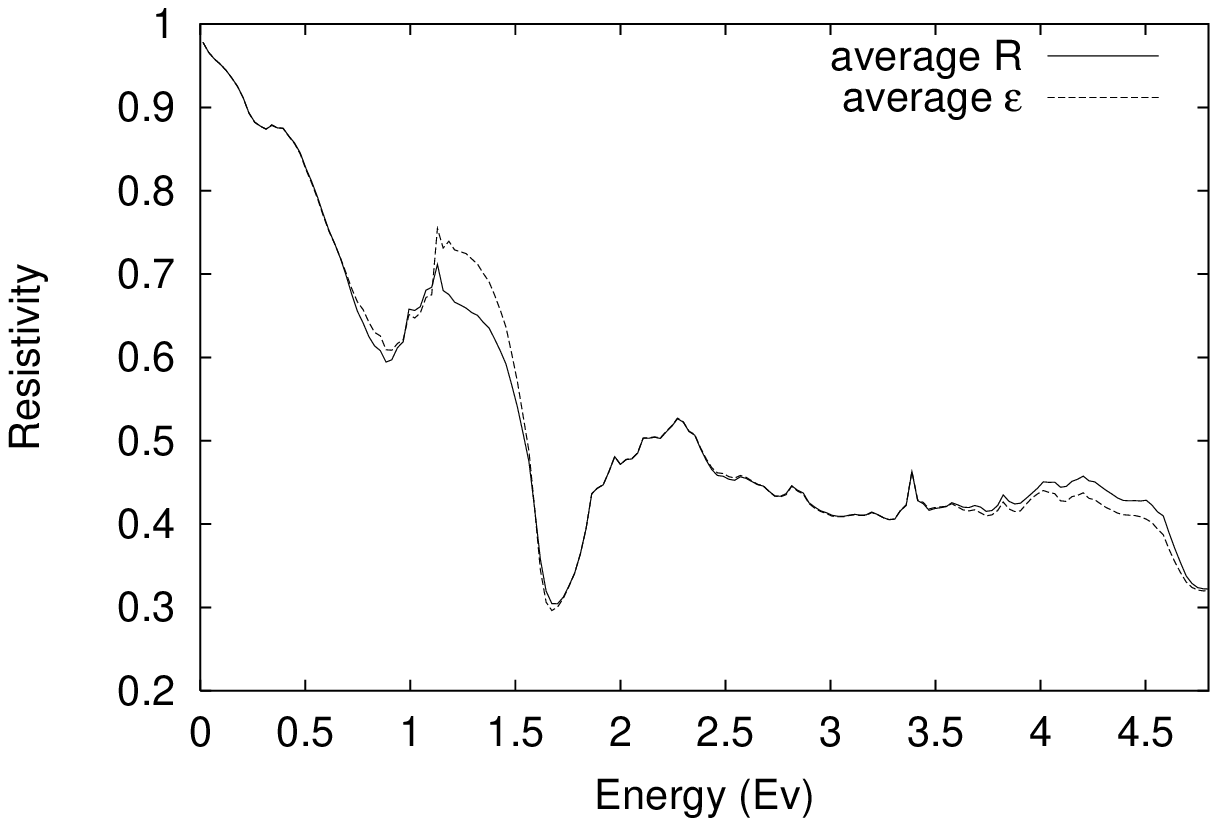,width=0.95\columnwidth}}
\vspace{0.1in}
\nopagebreak
\caption{Polarization average of the calculated resistivity, 
compared with the resistivity corresponding to the polarization
averaged calculated dielectric function. Averaging over polarization is
performed assuming equal weights of the in-plane and out-of-plane
polarizations.}
\label{R}
\end{figure}

%\begin{figure}[tbp] \centerline{\epsfig{file=NMbands.eps,width=0.95\columnwidth}}
%\vspace{0.1in} \nopagebreak \caption{ Calculated band structure of nonmagnetic
%CrO$_2$.  The Fermi energy is denoted by the dashed horizontal line at 0.}
%\label{NMbands} \end{figure}

%\begin{figure}[tbp] \centerline{\epsfig{file=NMDOS.eps,width=0.95\columnwidth}}
%\vspace{0.1in} \nopagebreak
%\caption{(a) DOS of nonmagnetic CrO$_2$ on a per formula unit basis. The Fermi
%energy is at 0. (b) Blowup of the DOS near E$_{F}$.} \label{NMDOS} \end{figure}

\end{multicols}
\begin{table}\caption{Spin-flip scattering threshold (distance from the highest occupied spin-up
state to the lowest unoccupied spin-down state) in eV. HL, VWN, vBH, LM, and
PW91 denote, respectively, exchange correlation potentials of
Hedin-Lundquist, Vosko-Wilk-Nussair, von Barth-Hedin, Langreth-Mehl, and
Perdew-Wang. The last two are GGA functionals.\label{thresh}}
\begin{tabular}{llllllll}
{\scriptsize
LAPW (HL)}&{\scriptsize
LAPW (PW91)}&{\scriptsize
LMTO (vBH)}&{\scriptsize
LMTO (VWN)}&{\scriptsize
LMTO (LM)}&{\scriptsize
LMTO(PW91)}&{\scriptsize
LMTO (VWN)$^a$}&{\scriptsize
LMTO(PW91)$^a$}\\ 
\tableline      
0.31 & 0.69 & 0.24 & 0.31 & 0.42 & 0.55&0.50&0.62 \\ 
\end{tabular}
$^a$Using setup from Ref.\protect\cite{anis}.
\end{table}


\begin{references}
\bibitem{note1}  In Ref.\onlinecite{fuji} a value and $U\approx 3.4$ eV was
deduced by comparing the distance between the maxima in photoemission and in
BIS spectra ($\approx 5$ eV) with the distance between the maxima in
occupied and unoccupied DOS in LSDA calculations of
Ref. \onlinecite{schwartz} ($%
\approx 1.6$ eV) . Since
the splitting between occupied and empty $d$-states is
mostly due to exchange and crystal field, this 
procedure is somewhat uncertain, however.
\bibitem{points} We use the standard tetragonal symmetry notations; the 
points $\Gamma$, Z, X, M, A and R have coordinates, in units of $(\pi/a,
\pi/a,\pi/c)$, (000),(001),(100),(110), (101) and (111), respectively.
\bibitem{note3} From LMTO calculations. We did not perform LAPW calculations
for the antiferromagnetic ordering.
\end{references}
\end{document}